\documentclass[journal]{IEEEtran}
\usepackage{graphicx}

\usepackage{textcomp}
\ifCLASSINFOpdf
\else
\fi
\usepackage{amsmath}

\usepackage{amssymb}
\usepackage{color,soul}
\usepackage{multirow}
\begin{document}
\title{Geometric Solution of Image Degradation by Diffraction in Lensless Sensing and Microscopy}
%\title{High Numerical Aperture Illumination for Non-Computational Lensless Sensing and Microscopy}
%
%

%

\author{Sanjeev~Kumar,~\IEEEmembership{}
        Manjunatha~Mahadevappa,~\IEEEmembership{Senior~Member,~IEEE,}
        and~Pranab~Kumar~Dutta,~\IEEEmembership{Member,~IEEE}% <-this % stops a space
\thanks{S. Kumar and M. Mahadevappa are with the School of Medical Science and Technology, Indian Institute of Technology Kharagpur, 721302, India, e-mail: sanjeevsmst@iitkgp.ac.in.}% <-this % stops a space
\thanks{P.K. Dutta is with the Department of Electrical Engineering, Indian Institute of Technology Kharagpur, 721302, India.}% <-this % stops a space

\thanks{Manuscript received date; revised date.}}

\markboth{Journal,~Vol, No.}%
{Kumar \MakeLowercase{\textit{et al.}}: Bare Demo of IEEEtran.cls for IEEE Journals}

\maketitle

\begin{abstract}
This paper proposes a non-computational method of counteracting the effect of image degradation introduced by the diffraction phenomenon in lensless microscopy. All the optical images (whether focused by lenses or not) are diffraction patterns, which preserve the visual information upto a certain extent determined by the size of the point spread functions, like airy disks in some cases. A highly diverging beam can be exploited to reduce the spatial extent of these point spread functions relatively in the transformed projective space, which can help us in the spatial unmixing of the visual information. The principle has been experimentally validated by the lensless imaging of red blood cells of diameter $\sim$6-9 $\mu$m and a photolithography mask with features in micrometer scale. The important advantages of the proposed approach of non-computational shadow microscopy are the improved depth of field and a drastic increase in the sensor to sample working distance. The imaging method can also be used as a projection technique in the multi-angle optical computed tomography (CT).
\end{abstract}


\IEEEpeerreviewmaketitle

\section{Introduction}
\begin{figure}
\centering
{\includegraphics[width=6cm]{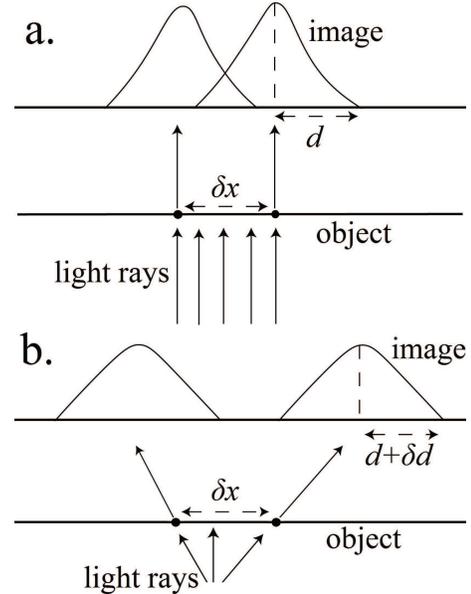}}
\caption{(a) A parallel beam illuminating two points and the corresponding image. (b) A diverging beam illuminating same two points and the corresponding image.}
\label{Fig:1}
\end{figure}
A lensless shadow microscopy system contains a light source to illuminate a weakly absorbing object and an image sensor to record this object\textquotesingle s shadow. Shadow microscopy with visible light has been discussed in the literature, for instance by \cite{yang2018resolution,fang2017chip,ozcan2016lensless} and \cite{wu2012optical}. The restriction to deposit sample directly on the sensor\textquotesingle s surface, for preserving the high resolution limited its use to on-chip cell or tissue cultures only. For instance, in the paper of Yang et al, 2018, the sample to sensor distance is 5 \ $\mu$m only, which is the thickness of the protective coating on the sensor surface \cite{yang2018resolution}. The fundamental problem that underlies lensless imaging is the scattering of the incident light by the sample and because of this phenomenon, every point in the object produces a cone of light which is projected as a point spread function (PSF) on the sensor. In incoherent illumination, this PSF can be approximated as a space-invariant Gaussian pattern whose variance depends on the sample to sensor distance \cite{muhammad2012sampling}. Because of the very narrow frequency bandwidth support of this Gaussian pattern, the reconstruction problem is severely ill-posed \cite{bertero2020introduction} and high resolution reconstruction of the object is very difficult (and has never been reported satisfactorily as per authors\textquotesingle \ literature survey). One solution is to use a coded mask in the beam path, so that the PSF takes the form of the magnified image of the mask itself (see \cite{wu2020single,cieslak2016coded,antipa2018diffusercam}). Now by the choice of a mask of larger frequency bandwidth support, a high resolution reconstruction of the object can be obtained using an appropriate algorithm. An alternate solution is to use a coherent illumination, where the PSF naturally has a very large frequency bandwidth support (unlike a Gaussian PSF). This approach is very common in lensless microscopy methods and is very well-known by the name of “lensless or digital in-line holographic microscopy” sometimes shortened as LIHM or DIHM (see \cite{ozcan2016lensless,kumar2020compressive,Kumar_2021,Kumar_2020} for details). Another similar lensless computational imaging technique used with the X-ray, ultraviolet light and electron beam is "coherent diffractive imaging" \cite{miao2011coherent,zurch2014real,clark2012high,yang2013coherent}.

\begin{figure}
\centering
{\includegraphics[width=8cm]{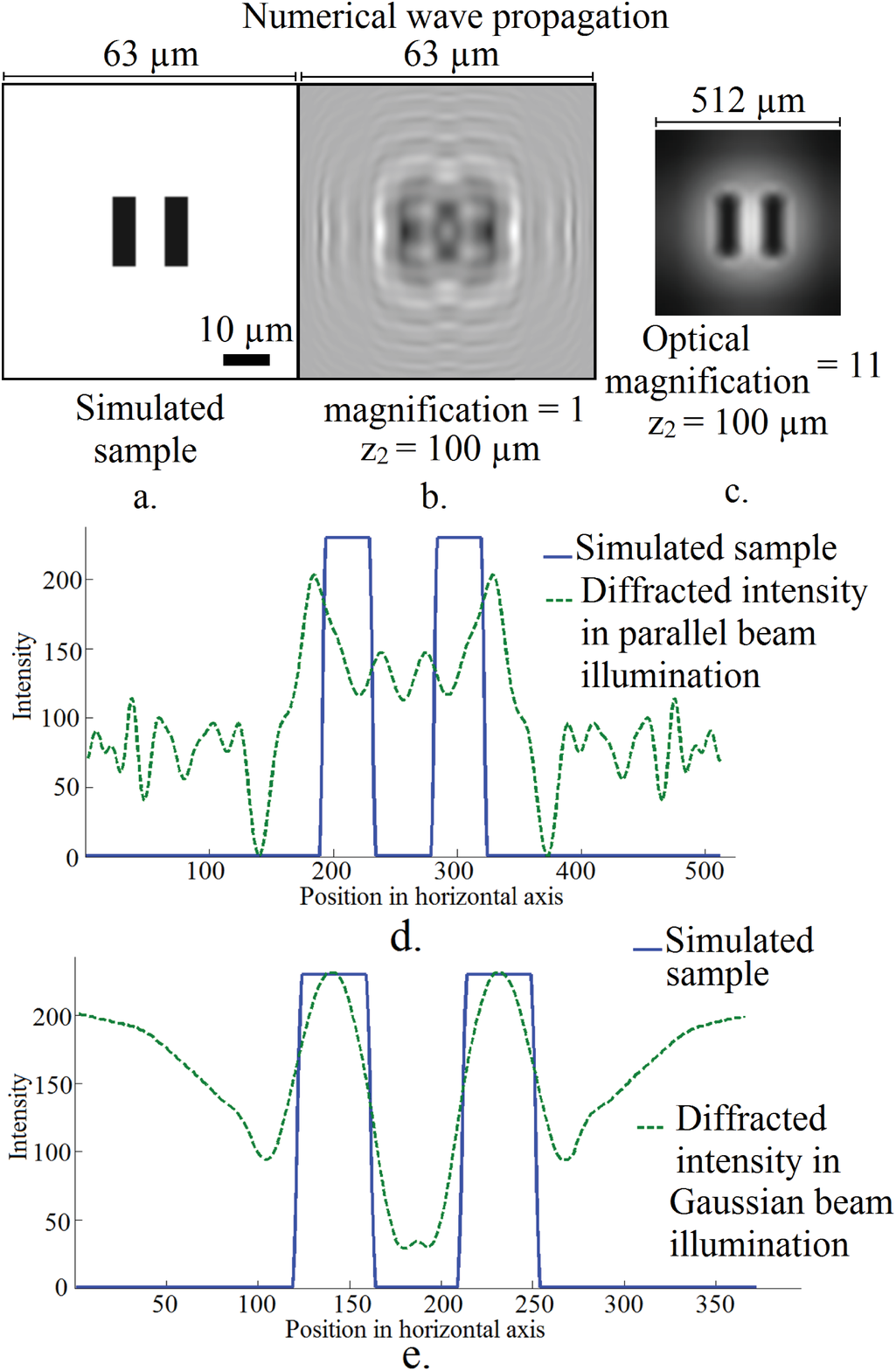}}
\caption{(a) A 512 $\times$ 512 sub-region from a 4096 $\times$ 4096 simulated test object. (b) Sub-region of the simulated diffraction image for the parallel beam illumination, sampled at pixel pitch of 0.125 $\mu$m, $z_2$ is the sample to sensor distance. (c) Simulated diffraction image for the Gaussian beam illumination corresponding to optical magnification value 11, downsampled to pixel pitch of 16 $\mu$m after numerical wave propagation. (d) Line plot of a single row (central row) from a and b. (e) Line plot of a single row (central row) from a and c, diffraction image is scaled appropriately before visualization to match the dimension of object in a.}
\label{Fig:1}
\end{figure}
\begin{figure*}
\centering
{\includegraphics[width=14cm]{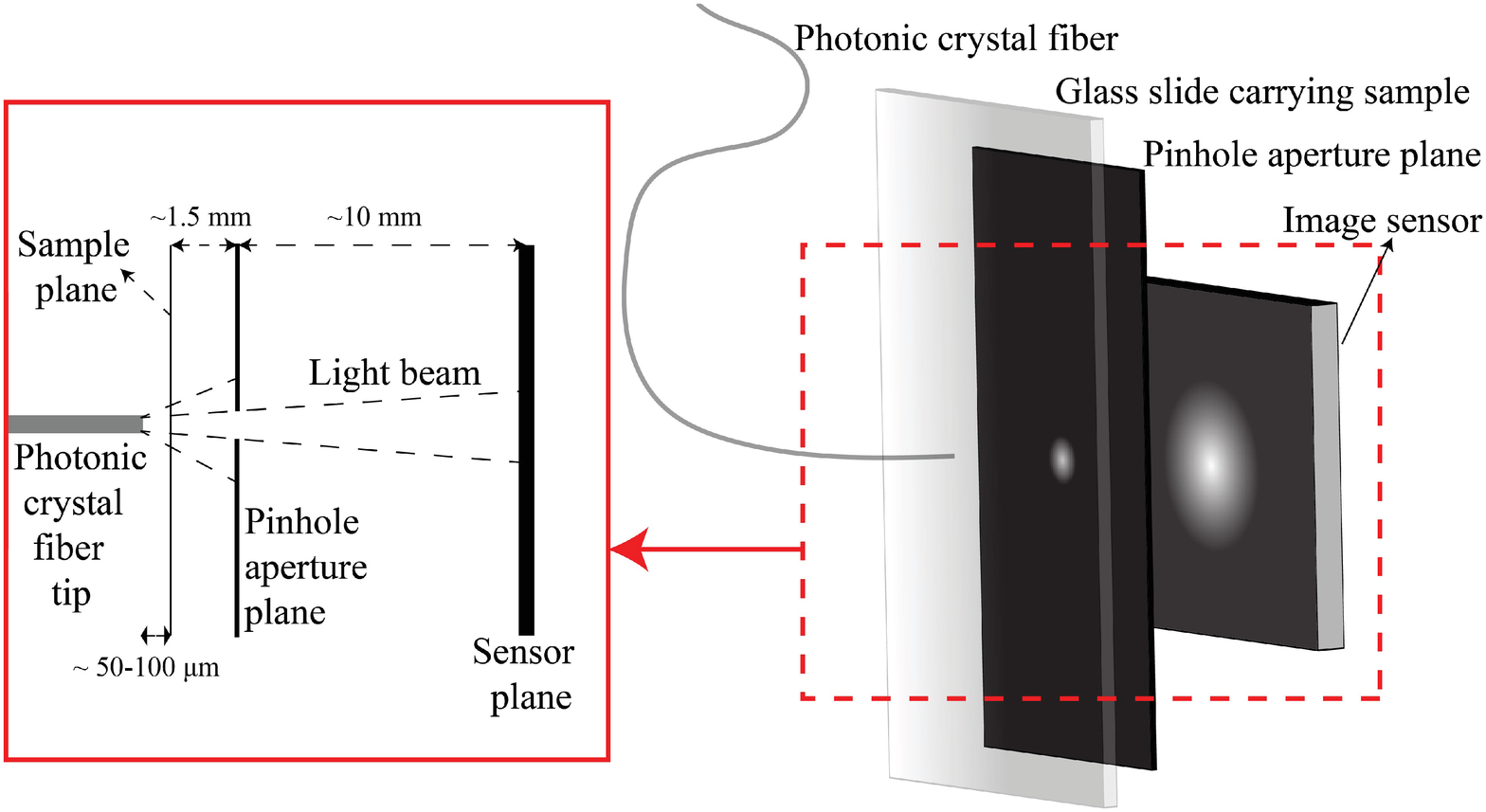}}
\caption{Imaging setup for the proposed method.}
\label{Fig:2}
\end{figure*}

\section{Principle}
Unlike coded aperture imaging and LIHM, this paper does not consider solving the inverse problem for reconstruction of the object from the measurements. Instead this paper focuses on a geometric solution for preserving the information about the microscopic sample. A diffraction pattern is an image of the sample (and vice versa \cite{mertz2019introduction}) which contains the visual information upto a certain resolution, determined by the spatial extent of the point spread function (PSF). Suppose there are two points in the object which are illuminated with a parallel beam of light. For a certain sample to sensor distance $z_2$, suppose the PSF has a radius $d$. We assume the spatial invariance of this PSF in our field of view for the sake of simplicity in this discussion. Now the signals from the individual points will start overlapping if the distance between these two points becomes less than $2d$ (see figure 1a). We can safely comment here that the half-pitch resolution limit is the radius of the PSF itself. (We are not considering the Rayleigh limit because the PSF here is an arbitrary pattern which will depend on the coherence of the light and the strength of unscattered part of the light. PSF shown in figure 1 is only indicative). Alternatively, if a high numerical aperture (NA) illumination i.e. a diverging beam is used to illuminate the same two points, a magnification $M$ will be introduced in the image. In this case, the half pitch resolution limit is not the radius of the PSF ($d+ \delta d$) but the ratio of ($d+ \delta d$) and $M$. In other words, the centers of the patterns from the two individual points are obtained at a larger separation, due to the different geometric projection angles for the different points (see figure 1b). The value of magnification is the ratio of distance between the light source and sensor plane ($z$) and the distance between the light source and sample/object plane ($z_1$) i.e. $M=\frac{z}{z_1}$ \cite{forsyth2002computer,hartley2003multiple}. This geometric magnification can be effectively exploited for non-computational shadow microscopy only if the magnification is large enough to counteract the loss of resolution due to the degradation introduced by the PSF.
\begin{figure}
\centering
{\includegraphics[width=8cm]{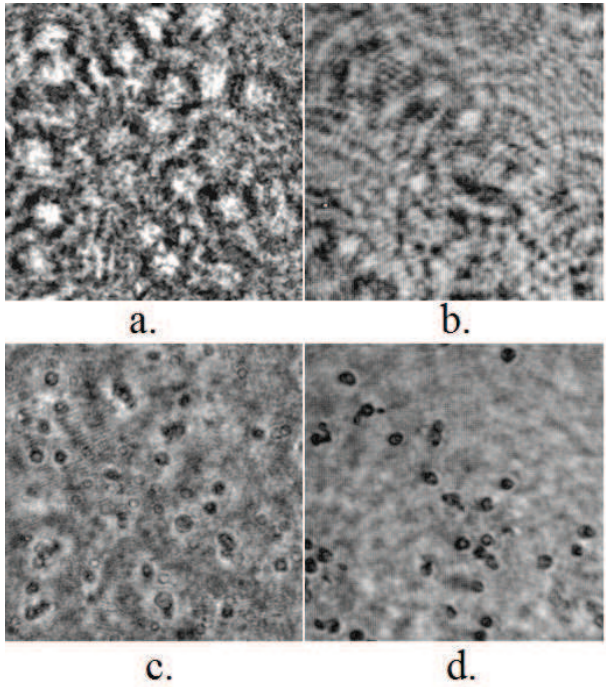}}
\caption{(a-b) Projections (diffraction patterns) of red blood cells (RBCs) of diameter $\sim$6-9 $\mu$m at magnification $\sim$1, wavelength 670 nm. (c-d) Corresponding reconstructions using the angular spectrum method. All the four images are of digital resolution 256$\times$256 and pixel-pitch 1.12 $\mu$m.}
\label{Fig:3}
\end{figure}

\begin{figure*}
\centering
{\includegraphics[width=16cm]{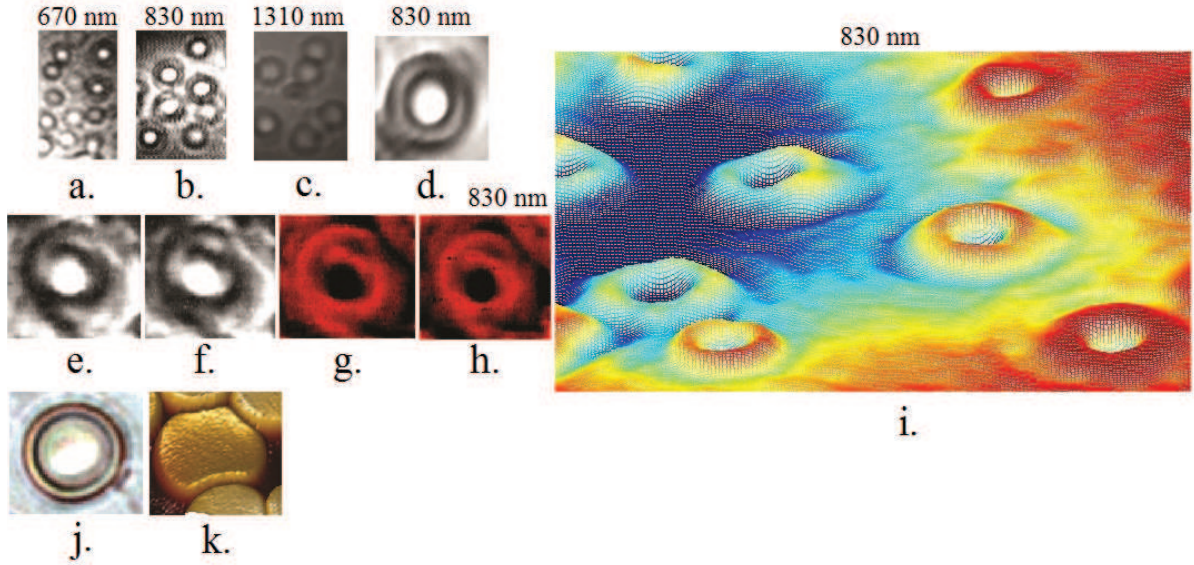}}
\caption{Projections of RBCs using the proposed method at different magnifications, pixel-pitch is $\sim$18 $\mu$m for all the images corresponding to the proposed method. (a-c) At magnification $\sim$100 under three different illumination wavelengths. (d)  At magnification $\sim$150, wavelength 830 nm. (e-f) At magnification $\sim$200, wavelength 830 nm. (g-h) At magnification $\sim$200 with pseudocolor, wavelength 830 nm. (i) At magnification $\sim$100 visualized using pseudocolor and mesh plot, blue denotes high intensity and red denotes low intensity, wavelength 830 nm. (j) Under bright-field microscope with 40x objective lens and broadband illumination. (k) Under atomic force microscope.}
\label{Fig:4}
\end{figure*}

\begin{figure}
\centering
{\includegraphics[width=6cm]{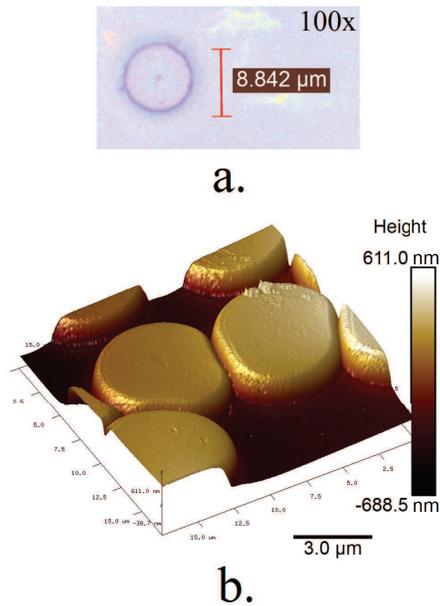}}
\caption{(a) RBC under a bright field microscope with 100x objective lens and broadband illumination. (b) RBCs under an atomic force microscope showing the height using color.}
\label{Fig:5}
\end{figure}
\begin{figure}
\centering
{\includegraphics[width=9cm]{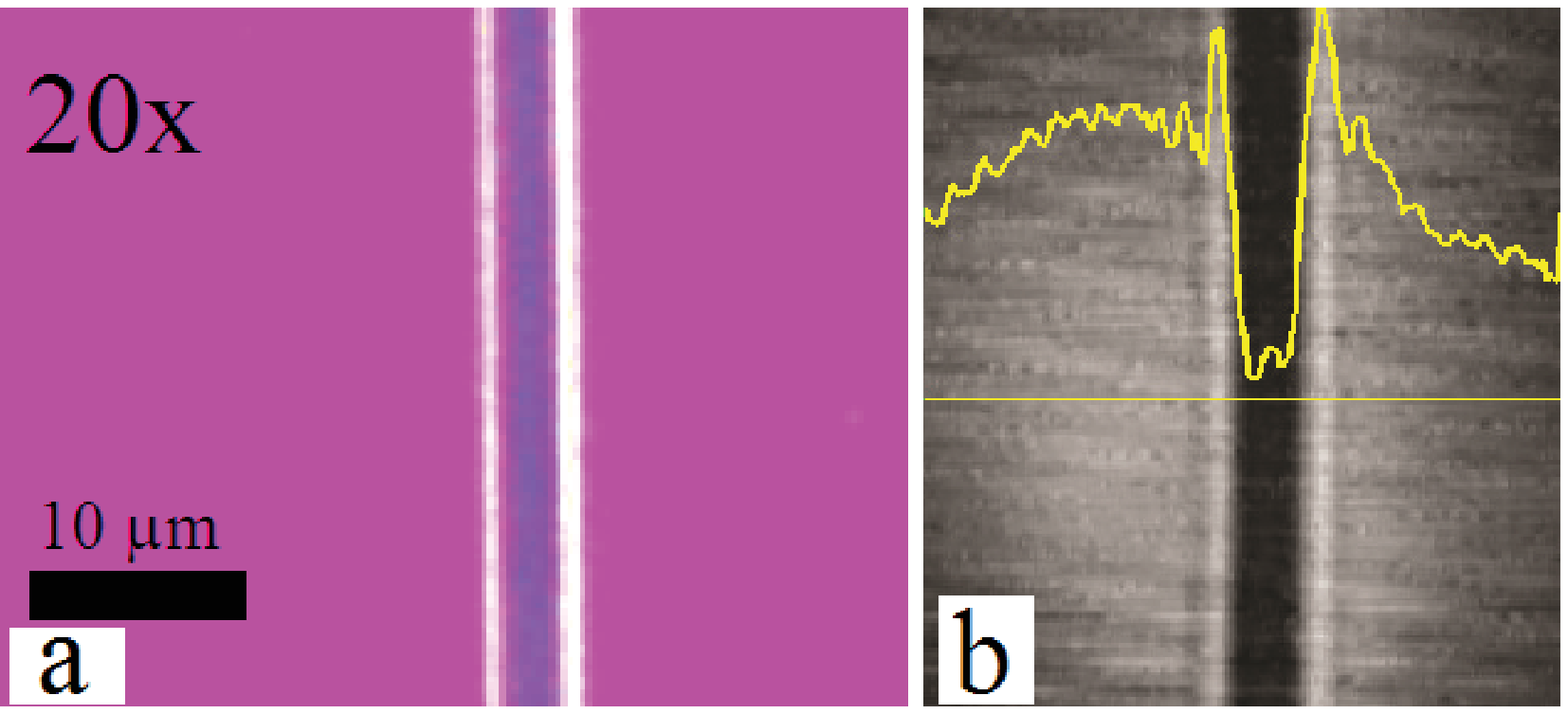}}
\caption{(a) A photolithography mask under a bright field microscope with 20x objective lens and broadband illumination. (b) Same mask\textquotesingle s image captured using the proposed method, digital resolution 178$\times$160, pixel-pitch $\sim$18 $\mu$m and wavelength 830 nm.}
\label{Fig:6}
\end{figure}

\section{Simulation experiments}
In this section, we validate the above described principle using the Fourier optics based simulation experiments.  For non-paraxial regime, the Rayleigh-Somerfeld (RS) diffraction integral or Angular spectrum (AS) method can be used for the numerical wave propagation \cite{goodman2005introduction}. AS method is computationally efficient and involves a point-wise multiplication with an appropriate optical transfer function in the frequency domain. The Fourier and the inverse Fourier transforms are obtained using the fast Fourier transform (FFT) and the inverse fast Fourier transform (IFFT) algorithms. The axial distance dependent optical transfer function $H(\textbf{v})$ can be obtained using the following equation \cite{kumar2020compressive,ozcan2016lensless}:
\begin{align}
H(\textbf{v}) &=  exp\Big(jk_0z_2 \sqrt{1 - (\lambda v_x)^2-(\lambda v_y)^2}\Big);  \sqrt{v^2_x+v^2_y} < \frac{1}{\lambda} \\
&= 0;\ \textrm{otherwise}
\end{align}
where $k_0 = \frac{2\pi}{\lambda}$ is the wave-number, $\lambda$ is the wavelength, $z_2$ is the distance, $(\textbf{v}=v_x,v_y)$ is the frequency coordinate vector.
Figure 2a shows a simulated test object with two bars of width 5 $\mu$m each. This image is a sub-region of a larger image of digital resolution 4096 $\times$ 4096 with a sampling distance of 125 nm. Figure 2b shows a sub-region of the simulated diffraction pattern for the case of parallel beam illumination, for a sample to sensor plane distance ($z_2$) of 100 $\mu$m. Figure 2b has the same digital resolution and spatial sampling as figure 2a.

In the second experiment, the same test object is multiplied with a simulated Gaussian beam (diverging beam illumination) and then numerically propagated to the same distance $z_2$ = 100 $\mu$m. The same digital resolution of 4096 $\times$ 4096 and spatial sampling of 125 nm has been maintained. After numerical propagation, the obtained diffraction pattern is downsampled to a sampling distance of the 16 $\mu$m, to mimic the low resolution imaging sensor used in the bench-top experiments performed in the following sections of this paper. Figure 2c shows the full field of view of the obtained diffraction image. The optical magnification value corresponding to the figure 2c is 11. As described in the previous section, the visual information about the bars which is lost in the case of the parallel beam illumination is well preserved when a highly diverging beam illumination is used, for the same sample to sensor plane distance.
This numerical experiment has been restricted to a small sample to sensor plane distance ($z_2$) because of the following computational limitations: since a high numerical aperture beam has been simulated, the object must be zero padded upto the appropriate lateral lengths before numerical propagation step. As the value of $z_2$ scales up, the memory requirement increases, reaching finally upto the computer capacity. Computational complexity will also scale up for the same reason. Breaking down the problem into the propagation-downsampling-propagation loop will introduce numerical (aliasing) errors related to the downsampling. Tomasz Kozacki et al, 2012 and Tomasz Kozacki, 2008 described these sampling related numerical errors in the angular spectrum method \cite{kozacki2012computation,kozacki2008numerical}.

\section{Materials and methods}
A solid core photonic crystal fiber (PCF) of numerical aperture (NA) $0.38 \pm 0.05$ and an effective mode field diameter of $1.8 \pm 0.3 \ \mu m$ (both the values at 780 nm respectively) has been used as an illumination source in this paper. This NA value corresponds to a half cone angle of 22.33 degrees. For a step index single mode optical fiber, the NA value is around 0.1, which corresponds to a half cone angle of 5.74 degrees only. The fiber is connected (using FC/PC connector) to a pigtailed laser diode of wavelength 830 nm and power 10 mW but the power output is manually controlled using a diode controller while imaging, to obtain the optimum contrast. Some images have also been captured with illumination wavelengths 670 nm and 1310 nm for the generalization. Sample fixed on a glass slide is mounted on a micrometer stage to control its three-dimensional position precisely. This provides a mechanical control of both the magnification and the lateral scanning of the sample. A pinhole aperture of arbitrary diameter of $<$ 1 mm has been kept between the sample and the sensor to select the rays from the features of interest (to some extent). This helps us to filter out any rays from the high scattering angles from the features outside the field of view of interest.

In lensless in-line holographic microscopy (LIHM) at unit magnification, the pixel-pitch of the image sensor determines the limit of resolution (along with some other factors), unless some sub-pixel super-resolution technique is employed. So a high resolution image sensor is an essential requirement in LIHM. In the proposed method, the high resolution of the image sensor (i.e. small pixel-pitch) is of little interest but a significantly large sensing area is the essential requirement. The reason can be understood with this instance: if a 25 $\mu$m feature is magnified by 200 times, it will be projected on a 5 mm sensor area. In this paper, a lead-oxysulfide vidicon image sensor of 9.5 mm (vertical) $\times$ 12.7 mm (horizontal) sensing area has been used. The horizontal and vertical resolutions of this image sensor is around 18 $\mu$m both (9.5 mm/525 vertical raster scans and  12.7 mm/700 horizontal TV lines). 

\section{Results and discussions}
To experimentally demonstrate the principle discussed in this paper, we first show two diffraction patterns captured at magnifications around 1 in the figures 4a and 4b. These figures show the projections of isolated red blood cells (RBCs) of diameter $\sim$6-9 $\mu$m. For these two figures, the spatial extent of the PSFs is much larger than the size of object under observation and hence no visual information can be obtained directly by looking at these projections. Reconstructions from these diffraction patterns (using angular spectrum method \cite{matsushima2009band} based on the principle of in-line holography) have been included in figure 4c and 4d, to assist the reader in recognizing the degradations introduced in these images by the diffraction phenomenon. Next we introduce a magnification of around 100 times using the optical setup and imaging geometry shown in figure 3 and the results have been shown in figures 5a to c. Unlike the figure 4, isolated RBCs can be observed easily in these images. This change in the visual appearance of the diffraction pattern and the spatial unmixing of the signals from individual RBCs in the latter images is the direct experimental validation of the principle presented in this paper. Figures 5e to h correspond to an optical magnification of around 200 times obtained using the same imaging geometry, now a single RBC image extends to a length of around $\sim$1.5-2 mm on the image sensor. For instance, for the figure 5e, the digital resolution is 96$\times$104 and pixel pitch is $\sim$18 $\mu$m (pixel pitch is same for all the images acquired with setup shown in figure 3). In these images, even the well-known concave shaped morphology of the RBCs \cite{merola2017tomographic} can be undoubtedly observed. The light source and the image sensor\textquotesingle s positions are fixed, only the sample\textquotesingle s position is changed to control the magnification. As the magnification is increased, the field of view decreases proportionally. Figure 6 shows the diameter and the thickness of an RBC measured in a bright field microscope and an atomic force microscope as the gold standard methods. 

In figure 7, images of a photolithography mask with features of dimensions $\sim$600 nm (first bright vertical line like feature from left), $\sim$3 $\mu$m (next dark vertical feature) and $\sim$1 $\mu$m (second bright vertical line like feature) have been shown for the further validation of the imaging principle and resolution. A full pitch resolution of around $\sim$2-3 micrometers can be anticipated for the proposed method, from the images of the red blood cells and this photolithography mask.

The images captured using this imaging geometry have one-point perspective, due to the depth dependent magnification. One advantage of this method is that the depth of field is not limited to few microns (or less than a micron) like in lens based microscopy operated at similar magnifications. Also the sensor to sample distance is not restricted to a few microns like the previous demonstrations of the lensless shadow microscopy \cite{yang2018resolution}, this eliminates the restriction to deposit sample on the sensor surface. The optical fiber tip was in a close proximity of the sample ($\sim$50-200 $\mu$m) which can be handled using an appropriate instrumentation. The stochastic vibration of the cleaved end of the optical fiber tip is also a factor contributing to the blurring of images. The proposed method is real-time and free of any numerical or aliasing error because of its non-computational nature. At last, the utility of this principle can be found in the design and development of cell counters, flow cytometry, imaging in microfluidics, crack detection and in the development of novel microscopy technologies. The imaging method can also be used as a projection technique in the multi-angle optical computed tomography (CT).

\section{Conclusions}
In conclusion, the image degradation introduced in the lensless shadow microscopy by the diffraction phenomenon can be suppressed by changing the illumination strategy. Geometric magnifications of large values obtained using a large numerical aperture light source, can help to us perform lensless imaging without involving any computational reconstruction step. Sub-pixel resolution has been clearly demonstrated, as the pixel-pitch of the image sensor used was $\sim$18 $\mu$m and the sample being captured was of size $\sim$6-9 $\mu$m or of further smaller size. Working distance increased drastically from few micrometers to few millimeters using the proposed principle.

\ifCLASSOPTIONcaptionsoff
  \newpage
\fi



%\bibliographystyle{IEEEtran}
\bibliographystyle{IEEEtran}
\bibliography{Sanjeev_NCLM}

\begin{thebibliography}{10}
\providecommand{\url}[1]{#1}
\csname url@samestyle\endcsname
\providecommand{\newblock}{\relax}
\providecommand{\bibinfo}[2]{#2}
\providecommand{\BIBentrySTDinterwordspacing}{\spaceskip=0pt\relax}
\providecommand{\BIBentryALTinterwordstretchfactor}{4}
\providecommand{\BIBentryALTinterwordspacing}{\spaceskip=\fontdimen2\font plus
\BIBentryALTinterwordstretchfactor\fontdimen3\font minus
  \fontdimen4\font\relax}
\providecommand{\BIBforeignlanguage}[2]{{%
\expandafter\ifx\csname l@#1\endcsname\relax
\typeout{** WARNING: IEEEtran.bst: No hyphenation pattern has been}%
\typeout{** loaded for the language `#1'. Using the pattern for}%
\typeout{** the default language instead.}%
\else
\language=\csname l@#1\endcsname
\fi
#2}}
\providecommand{\BIBdecl}{\relax}
\BIBdecl

\bibitem{yang2018resolution}
C.~Yang, H.~Ma, X.~Cao, X.~Hua, X.~Bu, L.~Zhang, T.~Yue, and F.~Yan,
  ``Resolution-enhanced lensless color shadow imaging microscopy based on large
  field-of-view submicron-pixel imaging sensors,'' in \emph{Proceedings of the
  IEEE Conference on Computer Vision and Pattern Recognition Workshops}, 2018,
  pp. 2246--2253.

\bibitem{fang2017chip}
Y.~Fang, N.~Yu, R.~Wang, and D.~Su, ``An on-chip instrument for white blood
  cells classification based on a lens-less shadow imaging technique,''
  \emph{PloS one}, vol.~12, no.~3, p. e0174580, 2017.

\bibitem{ozcan2016lensless}
A.~Ozcan and E.~McLeod, ``Lensless imaging and sensing,'' \emph{Annual review
  of biomedical engineering}, vol.~18, pp. 77--102, 2016.

\bibitem{wu2012optical}
J.~Wu, G.~Zheng, and L.~M. Lee, ``Optical imaging techniques in microfluidics
  and their applications,'' \emph{Lab on a Chip}, vol.~12, no.~19, pp.
  3566--3575, 2012.

\bibitem{muhammad2012sampling}
M.~Muhammad and T.-S. Choi, ``Sampling for shape from focus in optical
  microscopy,'' \emph{IEEE transactions on pattern analysis and machine
  intelligence}, vol.~34, no.~3, pp. 564--573, 2012.

\bibitem{bertero2020introduction}
M.~Bertero, \emph{Introduction to inverse problems in imaging}.\hskip 1em plus
  0.5em minus 0.4em\relax CRC press, 2020.

\bibitem{wu2020single}
J.~Wu, H.~Zhang, W.~Zhang, G.~Jin, L.~Cao, and G.~Barbastathis, ``Single-shot
  lensless imaging with fresnel zone aperture and incoherent illumination,''
  \emph{Light: Science \& Applications}, vol.~9, no.~1, pp. 1--11, 2020.

\bibitem{cieslak2016coded}
M.~J. Cie{\'s}lak, K.~A. Gamage, and R.~Glover, ``Coded-aperture imaging
  systems: Past, present and future development--a review,'' \emph{Radiation
  Measurements}, vol.~92, pp. 59--71, 2016.

\bibitem{antipa2018diffusercam}
N.~Antipa, G.~Kuo, R.~Heckel, B.~Mildenhall, E.~Bostan, R.~Ng, and L.~Waller,
  ``Diffusercam: lensless single-exposure 3d imaging,'' \emph{Optica}, vol.~5,
  no.~1, pp. 1--9, 2018.

\bibitem{kumar2020compressive}
S.~KUMAR, M.~Mahadevappa, and P.~K. Dutta, ``Compressive holography from
  poisson noise plagued holograms using expectation-maximization,'' \emph{IEEE
  Transactions on Computational Imaging}, 2020.

\bibitem{Kumar_2021}
\BIBentryALTinterwordspacing
S.~Kumar, M.~Mahadevappa, and P.~K. Dutta, ``Lensless in-line holographic
  microscopy with light source of low spatio-temporal coherence,'' \emph{IEEE
  Journal of Selected Topics in Quantum Electronics}, vol.~27, no.~4, p. 1–8,
  Jul 2021. [Online]. Available:
  \url{http://dx.doi.org/10.1109/JSTQE.2020.3028692}
\BIBentrySTDinterwordspacing

\bibitem{Kumar_2020}
\BIBentryALTinterwordspacing
------, ``Photonic crystal fiber for high resolution lensless in-line
  holographic microscopy,'' \emph{Optical Fiber Technology}, vol.~58, p.
  102248, Sep 2020. [Online]. Available:
  \url{http://dx.doi.org/10.1016/j.yofte.2020.102248}
\BIBentrySTDinterwordspacing

\bibitem{miao2011coherent}
J.~Miao, R.~L. Sandberg, and C.~Song, ``Coherent x-ray diffraction imaging,''
  \emph{IEEE Journal of selected topics in quantum electronics}, vol.~18,
  no.~1, pp. 399--410, 2011.

\bibitem{zurch2014real}
M.~Z{\"u}rch, J.~Rothhardt, S.~H{\"a}drich, S.~Demmler, M.~Krebs, J.~Limpert,
  A.~T{\"u}nnermann, A.~Guggenmos, U.~Kleineberg, and C.~Spielmann, ``Real-time
  and sub-wavelength ultrafast coherent diffraction imaging in the extreme
  ultraviolet,'' \emph{Scientific reports}, vol.~4, no.~1, pp. 1--5, 2014.

\bibitem{clark2012high}
J.~Clark, X.~Huang, R.~Harder, and I.~Robinson, ``High-resolution
  three-dimensional partially coherent diffraction imaging,'' \emph{Nature
  communications}, vol.~3, no.~1, pp. 1--6, 2012.

\bibitem{yang2013coherent}
W.~Yang, X.~Huang, R.~Harder, J.~N. Clark, I.~K. Robinson, and H.-k. Mao,
  ``Coherent diffraction imaging of nanoscale strain evolution in a single
  crystal under high pressure,'' \emph{Nature communications}, vol.~4, no.~1,
  pp. 1--6, 2013.

\bibitem{mertz2019introduction}
J.~Mertz, \emph{Introduction to optical microscopy}.\hskip 1em plus 0.5em minus
  0.4em\relax Cambridge University Press, 2019.

\bibitem{forsyth2002computer}
D.~A. Forsyth and J.~Ponce, \emph{Computer vision: a modern approach}.\hskip
  1em plus 0.5em minus 0.4em\relax Prentice Hall Professional Technical
  Reference, 2002.

\bibitem{hartley2003multiple}
R.~Hartley and A.~Zisserman, \emph{Multiple view geometry in computer
  vision}.\hskip 1em plus 0.5em minus 0.4em\relax Cambridge university press,
  2003.

\bibitem{goodman2005introduction}
J.~W. Goodman, \emph{Introduction to Fourier optics}.\hskip 1em plus 0.5em
  minus 0.4em\relax Roberts and Company Publishers, 2005.

\bibitem{kozacki2012computation}
T.~Kozacki, K.~Falaggis, and M.~Kujawinska, ``Computation of diffracted fields
  for the case of high numerical aperture using the angular spectrum method,''
  \emph{Applied optics}, vol.~51, no.~29, pp. 7080--7088, 2012.

\bibitem{kozacki2008numerical}
T.~Kozacki, ``Numerical errors of diffraction computing using plane wave
  spectrum decomposition,'' \emph{Optics Communications}, vol. 281, no.~17, pp.
  4219--4223, 2008.

\bibitem{matsushima2009band}
K.~Matsushima and T.~Shimobaba, ``Band-limited angular spectrum method for
  numerical simulation of free-space propagation in far and near fields,''
  \emph{Optics express}, vol.~17, no.~22, pp. 19\,662--19\,673, 2009.

\bibitem{merola2017tomographic}
F.~Merola, P.~Memmolo, L.~Miccio, R.~Savoia, M.~Mugnano, A.~Fontana,
  G.~D'ippolito, A.~Sardo, A.~Iolascon, A.~Gambale \emph{et~al.}, ``Tomographic
  flow cytometry by digital holography,'' \emph{Light: Science \&
  Applications}, vol.~6, no.~4, pp. e16\,241--e16\,241, 2017.

\end{thebibliography}

%\begin{thebibliography}{1}


%\end{thebibliography}


%\end{IEEEbiography}

%\begin{IEEEbiographynophoto}{Sanjeev Kumar}


%\newpage



%\vfill


\end{document}